\documentclass{vgtc}                          %

\graphicspath{{figures/}{pictures/}{images/}{./}} %

\usepackage{times}                     %

\usepackage{tabu}                      %
\usepackage{booktabs}                  %
\usepackage{lipsum}                    %
\usepackage{mwe}                       %
\usepackage{amsmath}
\usepackage{mathptmx}                  %
\usepackage{enumitem}
\usepackage{xcolor}
\usepackage{soul}
\usepackage{multirow}

\newcommand{\etal}{et al.}
\newcommand{\etals}{et al.'s}

\newcommand{\ie}{{i.e.,}}
\newcommand{\eg}{{e.g.,}}

\newcommand{\secref}[1]{\hyperref[#1]{Sec.~\ref*{#1}}}
\newcommand{\appendixref}[1]{\hyperref[#1]{Appendix~\ref*{#1}}}
\newcommand{\figref}[1]{\hyperref[#1]{Fig.~\ref*{#1}}}
\newcommand{\eqnref}[1]{\hyperref[#1]{Eqn.~\ref*{#1}}}
\newcommand{\tabref}[1]{\hyperref[#1]{Table ~\ref*{#1}}}

\definecolor{quoteColor}{HTML}{ff5733}

\newcommand{\parahead}[1]
{%
  \paraheadd{#1}.
}

\newcommand{\paraheadd}[1]
{%
  \vspace{0.07in}%
  \noindent%
  \textbf{\textit{#1}}%
}

\def\subsubsec#1
{\subsubsection{#1}}

\onlineid{7140}

\vgtccategory{Research}

\vgtcinsertpkg

\title{Ten Years Later: Replicating Two Color Discrimination Studies

}
\usepackage{orcidlink}

\author{Shadmaan Hye \orcidlink{0000-0002-9713-8457}\\ %
        \scriptsize University of Utah %
\and Andrew M. McNutt \orcidlink{0000-0001-8255-4258}
\\ %
     \scriptsize University of Utah %
\and Katherine E. Isaacs \orcidlink{0000-0002-9947-928X}
\\ %
     \parbox{1in}{\scriptsize \centering University of Utah}}

\abstract{
Color discrimination is a fundamental aspect of visualization as it influences how people interpret visual encodings. 
Many visualization guidelines are informed by perceptual studies, yet relatively few have been replicated.
Acknowledging that the interaction between human perception, visual tasks, and display technology can change over time, we replicate two crowdsourced color discrimination studies conducted 10 years earlier.  
Specifically, we replicated a visualization-focused color discrimination task ($N=144$) and a more general perceptual discrimination task ($N=394$). 
In both studies, our results reproduced the original perceptual effects.
We further use the replication to investigate whether color-related practice influences color discrimination. 
Specifically, we extended our replication studies by adding questions about participants' engagement with color practices. We then examined whether diverse color-related practices (\eg{} artistic hobbies, knowledge of color theory, and cosmetic makeup use) influenced color discrimination. We found no significant difference between participants who reported engaging in color-related practices and those who did not, suggesting that design guidance regarding color discrimination may generalize across viewers regardless of their regular color practice.

}
\keywords{Color Perception, Replication, User Study}

\begin{document}

\maketitle

\section{Introduction}
\label{sec:intro}

Color discriminability is essential for visualization, since the amount of visual information a color encoding can convey depends on the viewer's ability to distinguish among the colors used to encode data. 
Accordingly, prior work has investigated factors that shape color discriminability, including lightness variation, color axes, and stimulus properties~\cite{szafir2018modelingColorDifferenceF, reinecke2016enablingDesignersToFores, stone2014engineering, ware2023rainbow, LaceMariah2017HUeBandsandPerception}. 
Practitioners use these findings to inform visualization design guidelines and color palette tools~\cite{GramazioKaren2017CreatingColorPalettesforInfoVis, Bartram2011WhisperDontSceram, heer2012color, Borland2007rainbow}.
Recognizing the importance of these findings, we replicated two influential color-discrimination studies: Szafir's~\cite{szafir2018modelingColorDifferenceF} study of color discrimination with scatterplots, and Reinecke~\etals~\cite{reinecke2016enablingDesignersToFores} study of general color-discrimination.

Our Experiment~1 (\secref{sec:exp1}) replicates the scatterplot study, which quantified perceptual color differences across varying mark sizes ($N=144$).
Our Experiment~2 (\secref{sec:exp2}) replicates the just-noticeable difference study between an oriented shape and its background ($N=394$). 
We preserve the original tasks and analyses while reimplementing the studies on a new platform (reVISit~\cite{cutler2026revisit}) with a new participant pool (Prolific~\cite{PALAN201822}).
Across both experiments, we reproduced the original perceptual effects: color discrimination improved as color differences and contrast increased, closely matching the original findings. These results provide additional evidence that these perceptual effects remain robust across implementations, participant samples, and experimental platforms.

In addition to validating original study findings, replication provides a basis for asking new questions~\cite{Quadri2019CantPublishReplication}.
We use this opportunity to investigate whether color practice affects color discrimination. 
Perceptual-learning studies show that repeated color-discrimination training can improve performance, but these improvements are typically specific to the trained stimuli~\cite{horiuchi2024color}. 
It remains unclear whether the repeated color comparisons encountered in other color-related activities also improve performance. 
In each replication, we included a questionnaire about color-related practices (\eg{} artistic hobbies, color-theory knowledge, and cosmetic makeup use), which we used to partition participants into those who reported such practices and those who did not.
In both experiments, we found no significant differences in color discrimination between participants with and without self-reported color-related practices. 
This finding suggests that reported color practice does \emph{not} affect color discrimination in a manner measurable through these tasks. 
We see this as a positive (null) result, as it indicates that this key component of visualization does not need to change to accommodate differences in its designers' or audiences' experience with color-related activities.

This paper contributes
(1) confirmatory \textbf{replications} of two color-discrimination studies 
and (2) \textbf{extension} of those studies showing no measurable effect of color practice on color discrimination.
Both studies were marked exempt by our institution's IRB. 

\section{Related Works}
\label{sec:literature}

We draw on prior work on color discriminability and experience.

\parahead{Color discriminability in visualization}
Color is effective as a visual encoding only when viewers can reliably distinguish the colors used to represent data. Early visualization work treated color selection as a perceptual design problem, considering factors such as color distance, linear separation, and color category when choosing colors for data visualization~\cite{healey1996choosing}. However, standard color-difference models such as CIELAB can misestimate perceived differences when colors appear in realistic visualization contexts rather than as large, isolated patches under controlled viewing conditions~\cite{szafir2014adapting, stone2014engineering}. Subsequent work has measured how color discriminability varies with brightness, contrast, mark size, mark type, color distance, and display conditions~\cite{stone2014engineering, szafir2018modelingColorDifferenceF,reinecke2016enablingDesignersToFores}. These findings inform practical color-design systems and empirical evaluations, including tools for discriminable palette construction~\cite{GramazioKaren2017CreatingColorPalettesforInfoVis}, evaluations of quantitative colormaps~\cite{liu2018somewhere}, and models of feature-detection thresholds in colormaps~\cite{ware2019measuring}. Our work contributes to this line of research by replicating two studies that measured color-discrimination thresholds relevant to visualization design.

\parahead{Color-related experience and perceptual learning}
Our extension is motivated by evidence that perceptual performance can vary across observers and can change with experience. Prior work has reported population-level differences in color naming and categorization~\cite{mylonas2014gender, fider2019differences}, and contextual conditions can influence color judgments~\cite{Kobayashi2025SeeingIsntBelieving}. Perceptual-learning studies further show that repeated color-discrimination training can improve performance, although these gains are often specific to trained hues, locations, or stimulus conditions~\cite{ozgen2002acquisition, horiuchi2024color}. More broadly, expertise and domain training can affect visual processing strategies and task performance~\cite{Hall2022ProfessionalDifferences, robson2021EffectOfExpertise}. However, it remains unclear whether everyday color-related activities, such as artistic hobbies, knowledge of color theory, or cosmetic color matching, transfer to the color-discrimination tasks used in visualization. We address this question by extending both replications to include questionnaires on participants' color-related practices.

\section{Experiment One: Discrimination in Scatterplots}
\label{sec:exp1}

We replicate Szafir's~\cite{szafir2018modelingColorDifferenceF} study of color discrimination, which models perceptual color differences across varying point sizes. Among the original study's three experiments, we specifically focus on the scatterplot experiment, which directly measures the relationship between mark size and color discriminability.

\begin{figure}[t]
  \centering
  \includegraphics[width=\linewidth]{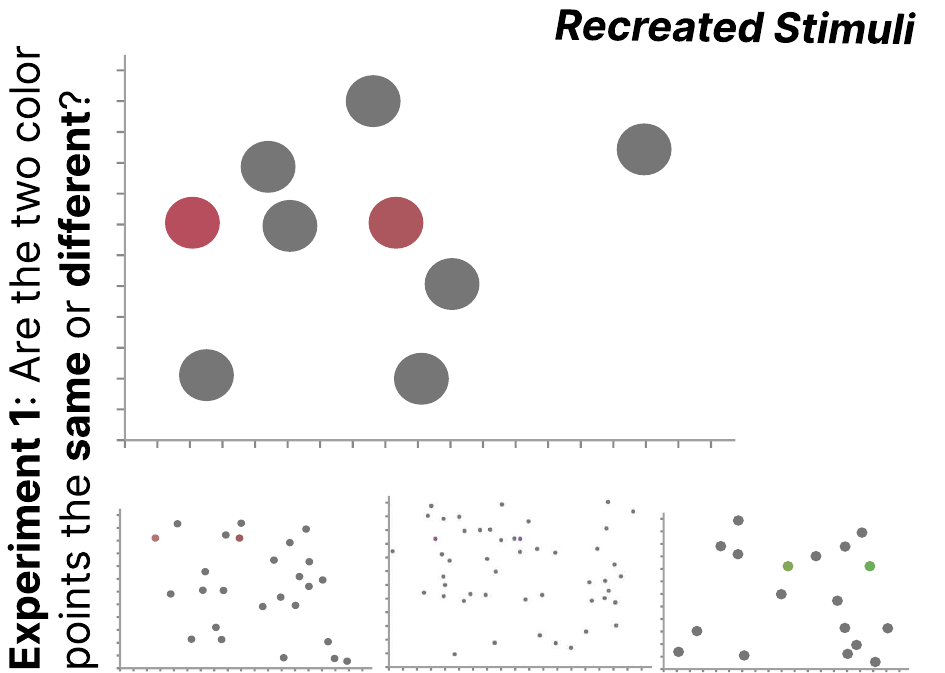}
  \caption{
  Stimuli used in Experiment 1.
  }
  \label{fig:exp_1_stimuli}
  \vspace{-2em}
\end{figure}

\parahead{Stimuli}
We recreated the stimuli as described in the original paper. Each stimulus consisted of two colored target marks placed in a field of mid-gray distractor marks designed to represent the visual clutter, yielding a scatterplot as in \figref{fig:exp_1_stimuli}. The stimuli were rendered in a white background with minimal gray axes. Mark diameters ranged from 0.25° to 2.0° of visual angle, and the two target marks were separated by 5° of visual angle, or roughly 125 pixels, to maintain consistent spatial comparison across stimuli.
Test points were assigned uniform random y-values. Distractor marks points were calculated via: $\sqrt{\text{visualization area}/\text{mark area}}$ with locations sampled from a normal distribution. 
We use the color set from the original study, which consists of 79 colors, uniformly sampled within the CIELAB gamut, spanning three axes: L*, a*, and b*.
For each target color, a comparison color was created by modifying only a single color axis while keeping the other two constant. For example, while testing the lightness L* axis, the values of axes a* and b* were kept constant, whereas the L* value was adjusted by a predetermined color difference amount, $\Delta E$, prescribed in the original study (in this experiment, we measure $\Delta E$ as Euclidean distance in LAB space). This gives 72 color pairs for each axis. 

\parahead{Procedure}
Participants provided informed consent and reviewed a brief overview of the procedure. After that, they completed a color vision screening using four Ishihara plates~\cite{ishihara1951tests}. 
They then finished a tutorial with three example stimuli showing identical colors, hue differences, and lightness differences, respectively, and had to answer the tutorial questions correctly before continuing.
In the main experiment, participants completed a binary forced-choice task in which they judged whether two target marks were the same color or different. 
A mid-gray screen was shown between trials to reduce contrast effects.
Three large-difference and four identical-color stimuli were used as engagement checks.

Afterward, participants completed a short questionnaire. We asked about hobbies that typically involve considerations of color
(\eg{} painting, drawing, graphic design, photography), whether they had formal education in color theory, and two simple color-mixing questions to measure their knowledge of color. Finally, we included questions about familiarity and use of cosmetic makeup, since choosing among similar shades of cosmetics may induce a color discrimination task. 
These questions helped us explore whether participants' real-world color-matching practice influenced task performance. We also asked demographic questions regarding gender, age group, and highest level of education.
See \href{https://color-perception-replication.netlify.app/}{color-perception-replication.netlify.app} for the studies.

\parahead{Participants}
We recruited ($N=144$) participants from Prolific using built-in color-blindness screening and by limiting participants to the United States. 
We selected sample size ($N=144$) by doubling the number of participants from the original study, which had 24 participants per axis (L*, a*, b*), to account for the color practice conditions. 
Participants were grouped as \emph{Known Color Practice} (n = 93) and \emph{Unknown Color Practice} (n = 51) based on the post-study questionnaires. Participants with at least one form of color practice (artistic hobbies, color theory knowledge, or cosmetic makeup use) were classified as \emph{Known Color Practice}. The resulting groups were imbalanced because participants were recruited from the general Prolific population and classified after data collection. 

The majority of the participants fell in the 45–54 (27.8\%) and 35–44 (27.1\%) age groups. 
Most participants had a bachelor’s degree (43.8\%) or a high school diploma (33.3\%).
Participants were recruited at $\sim50$\% female and $\sim50$\% male to ensure a balanced sample across demographic groups.
This decision was informed by a pilot study with 10 participants across gender groups, which indicated variation in familiarity with color theory and makeup. 
Further, the study group partitioning was influenced by popular discussion~\cite{mylonas2014gender} of gender differences in color naming. These patterns do not generalize to all individuals, which is why we use them as a proxy rather than a direct signal. 

\parahead{Results} 
We successfully replicated the principal findings reported by Szafir~\cite{szafir2018modelingColorDifferenceF}. As in the original study, participants were better at discriminating colored marks as the color difference between the target marks increased, and performance improved with larger mark sizes. We then examined whether self-reported color-related practices influenced color discrimination and found no significant differences between groups, indicating that color practice \emph{does not} influence color discrimination ability.

\begin{figure}[t]
  \centering
  \includegraphics[width=\linewidth]{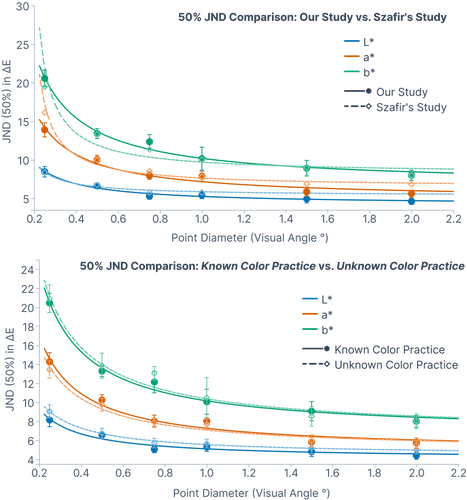} %
  \caption{
  Our replication of Szafir's color discrimination for scatterplots closely matched the original study results (top) and showed negligible differences between \emph{Known Color Practice} and \emph{Unknown Color Practice} participants (bottom). Study results are shown along the three CIELAB axes (L*, a*, b*), modeled as 50\% JNDs.
  }
  \label{fig:exp-1-results}
  \vspace{-2em}
\end{figure}

We summarize and compare our results with Szafir's in \figref{fig:exp-1-results}.
First, we grouped the different color trials by axis (L*, a*, b*), mark point size (0.25$^{\circ}$, 0.5$^{\circ}$, 0.75$^{\circ}$, 1$^{\circ}$, 1.5$^{\circ}$, and 2$^{\circ}$) and absolute value of $\Delta E$. Within each group, discriminability was measured as the proportion of correct responses. We then fit the original model, $p = m_x * \Delta x$, where $p$ is the proportion of detected color differences ($p = 50\%$ is a typical JND), $m_x$ is the regression line slope for the CIELAB axis, $x$, and $x \in \{L*, a*,b*\}$. For each axis-size combination, we converted the fitted slope into a 50\% threshold using $ND(50\%, s) = 0.5/m_x$, where $s$ is the mark size.  

Echoing the original study, within each axis, the threshold generally decreased as point diameter increased, indicating that larger marks made small color differences easier to distinguish. 
This pattern closely resembles the perceptual asymmetry reported in the original study, in which variations in L* were often easier to distinguish than those in a* or b* (see upper chart in \figref{fig:exp-1-results}, where lower value in vertical axis represents greater precision). Performance was also affected by mark size, with larger marks (further right on \figref{fig:exp-1-results}'s horizontal axes) leading to higher discriminability. 
As the point diameter increased, the threshold decreased across all channels, indicating improved sensitivity to color differences. For instance, at 0.25$^\circ$, thresholds for L*, a*, and b* were 8.46, 13.89, and 20.56, respectively, whereas at 2.0$^\circ$, they decreased to 4.61, 5.62, and 8.00, respectively. This reduction indicates that increasing mark size improves the ability to discriminate color differences.

We next partitioned participants based on their self-reported color-related practices, applied the same analysis to each group, and found no significant differences between the groups.
In particular, a Welch's two-sample t-test~\cite{west2021best} on participant-level accuracy showed no significant difference ($p=0.97$), indicating no observed effect of experience on color discrimination.
This null result suggests that basic use of color palettes in visualization can be used by anyone with non-atypical vision, regardless of color-practice experience. 

\parahead{Replication Differences} The overall trends match Szafir's, but the pattern is not uniform across conditions (\figref{fig:exp-1-results}, top). Larger mark sizes show the closest convergence, whereas in the smallest mark sizes (0.25$^{\circ}$), our thresholds are noisier (see Supplemental Material Table 1 for a quantitative comparison). Paired effect sizes (Cohen's $d$) show the largest differences on the L* ($d_z=-1.01$) and a* ($d_z=-0.95$) axes, with a smaller effect on b* ($d_z=0.72$). 
Because our study was conducted under different experimental conditions from the original, we cannot attribute these differences to any single factor.
One possible explanation is that on many participant screens, a mark size of 0.25$^{\circ}$ covers only a few pixels, making the stimulus more sensitive to differences in viewing setup and display configuration. 
The largest discrepancies between the original study and our study were observed on the L* axis. Notably, the original study was conducted nearly 10 years earlier, and in recent times, adaptive viewing technologies such as automatic brightness adjustment, HDR processing, and dynamic contrast have evolved substantially. Moreover, display hardware, rendering technologies, and viewing configurations have also evolved. Because the L* axis primarily reflects luminance differences, these technological advances may have contributed to the discrepancies. 
Differences in the experimental platform (reVISit~\cite{cutler2026revisit}), participant demographics, and other uncontrolled aspects of data collection may also have contributed.

\parahead{Limitations}
The study was limited to controlled online discrimination tasks, which isolate perceptual sensitivity. Because the experiments were conducted remotely, the display settings, calibration, and resolutions were not controlled. 
We recruited participants via Prolific and used their self-reported responses regarding color-related practices. These categories 
may leave out others with alternate forms of color practice. 
Due to limited access to original data and materials, we are unable to fully reanalyze the data from the original study in conjunction with ours---limiting the scope of analyses we could conduct.
Lastly, we replicated only one of the three studies considered in the original work, which also investigated bar charts and line charts. While our results largely agree with that work, they may differ for alternative stimuli. 

\begin{figure}[t]
  \centering
  \includegraphics[width=\linewidth]{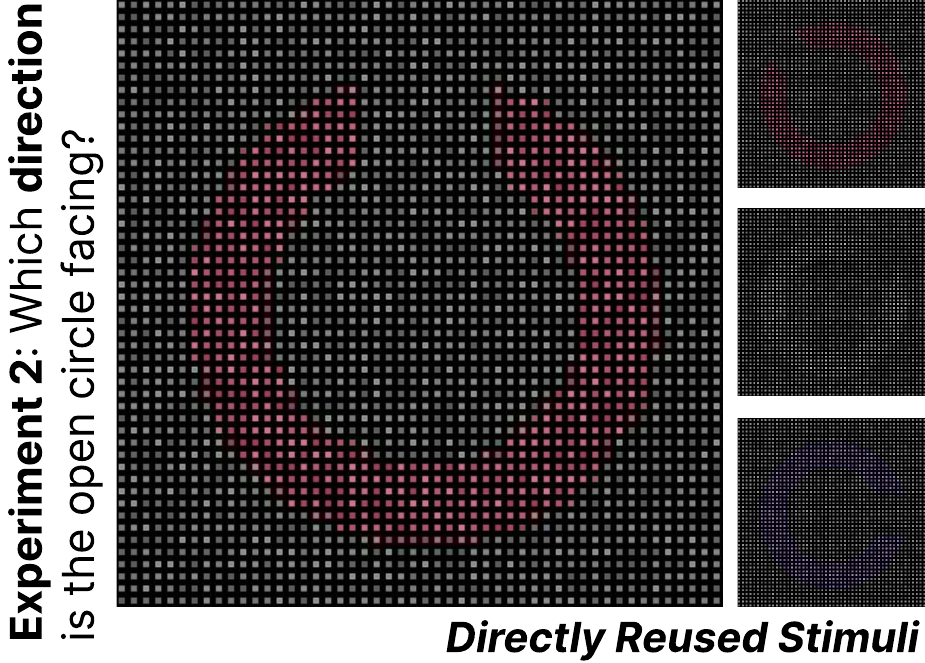}
  \caption{
  Stimuli used in Experiment 2.
  }
  \label{fig:exp_2_stimuli}
  \vspace{-2em}
\end{figure}

\section{Experiment Two: Discrimination in Landolt-C}
\label{sec:exp2}

Next, we replicated Reinecke \etal{}'s~\cite{reinecke2016enablingDesignersToFores} web-based color differentiation test. 
This experiment investigates a more general perceptual task, relating to the identification of visual orientation. 

\parahead{Stimuli}
The exact stimuli were obtained from the original study's authors. Each image contained a colored Landolt-C~\cite{ISO8596_2017} embedded in the center of a regular 50x50 cell grid. There were four color families in the data: red, blue, magenta, and lightness/white.

\parahead{Procedure}
Participants first reviewed an introduction and consent page, then completed a set of practice trials. 
The main experiment involved identifying the direction of a Landolt-C, which could take one of 8 cardinal directions. 
After each response, an iterated binary search adjusted the stimulus Landolt-C's color in discrete steps, moving it toward either the background color or full saturation until it converged on the participants' discrimination threshold. Thresholds are expressed as the number of steps from the background color, where a larger step count indicates that a greater color difference is needed to identify the Landolt-C's orientation.
Participants completed 12 sets, separated into three subblocks, in which the four different possible colors were randomly sorted. 
A short halfway break was included as an attention check midway through the experiment, after which participants completed the remaining sets. Following the task, participants completed a standard demographic questionnaire and two questionnaires (from Exp.~1) on color-related hobbies, color-theory knowledge, and familiarity with cosmetics. 
See \href{https://color-perception-replication.netlify.app/color-vision-perception}{https://color-perception-replication.netlify.app/color-vision-perception} for the study.

\parahead{Participants}
Using Prolific, we recruited 400 participants, comprising 200 males and 200 females, using the same screening as in Experiment~1.
According to the power analysis, approximately 375 participants were required to achieve 85\% power ($\alpha=.05$) to detect the target effect size. So we recruited 400 to account for participant exclusions and data loss.
Six participants (3 male and 3 female) were omitted due to data issues, resulting in a final sample of 394 participants.
Again, using post-study questionnaires (color-related hobbies, color theory knowledge, and cosmetic makeup usage), we grouped participants into \emph{Known Color Practice} (n = 252) and \emph{Unknown Color Practice} (n = 142).
The majority of participants were 25–34 (27.4\%) or 35–44 (25.9\%) years old. Most participants had a bachelor’s degree (39.3\%) or a high school diploma (36.8\%). 

\parahead{Results} 
Our results generally replicated the original results. As in Reinecke~\etal{}, we averaged each participant's thresholds across repetitions for each color, then reconstructed a discrimination ellipsoid in CIE L*u*v* color space (the region in 3D color space within which colors are predicted to be difficult for that participant to distinguish). Larger volumes indicate poorer color discrimination. For participant whose threshold combinations did not define a valid chromatic ellipse, we implemented a fitted fallback analysis, which fits a deterministic positive-definite ellipse to the observed red, magenta, and blue thresholds and combines it with the lightness threshold to compute ellipsoid volume. This analysis retained all participants and led to qualitative conclusions similar to those of the original study.
Finally, as in Experiment 1, our study showed no difference between those with and without color-related practices.

\begin{figure}[t]
  \centering
  \includegraphics[width=\linewidth]{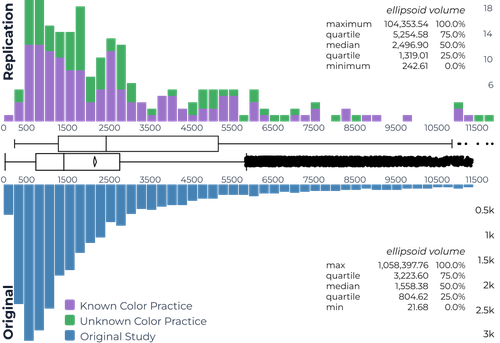}
  \caption{Our replication of Reinecke \etals{}~\cite{reinecke2016enablingDesignersToFores} perceptual direction task broadly follows the distribution found in the original study. 
  Both color practice groups had comparable distributions, indicating consistent perceptual patterns among participants.} 
  \label{fig:exp2-results}
  \vspace{-2em}
\end{figure}

Our replication reproduced the key results reported by Reinecke \etal{}~\cite{reinecke2016enablingDesignersToFores}. 
As in the original study, the reconstructed ellipsoid-volume distribution was positively skewed (\figref{fig:exp2-results}). Most participants ($\sim$63\%) produced relatively small discrimination volumes (\ie{} higher sensitivity), while a subset ($\sim$37\%) produced larger volumes, indicating lower sensitivity. 
We are unable to compare the distributions directly because the results from the original study are not available beyond summary statistics. 
Unlike the original study, we did not record participants' ambient lighting because our focus was on replicating the color-discrimination task, not on evaluating the effects of viewing conditions. 
As a result, we cannot account for variability in viewing environments, which may have introduced additional noise into the performance results.

We found no substantial difference between the \emph{Known Color Practice} and  \emph{Unknown Color Practice} groups. 
We assumed that accuracy was normally distributed in both groups, based on Shapiro-Wilk normality tests~\cite{razali2011power} that did not indicate otherwise. 
Further, Welch's t-tests revealed no statistically significant differences between the two groups. The trial accuracy did not differ between groups ($p= 0.674$).
Comparing the two groups in \figref{fig:exp2-results}, we can observe that the difference was not significant.

\parahead{Replication Differences} 
The replication reproduced the overall shape of the original discrimination-ellipsoid volume distribution, with both experiments showing stronger right skew. However, our participants' median ellipsoid volume was larger than that reported in the original study, suggesting that they often required larger color differences to discriminate colors.
These results suggest that the original perceptual pattern was successfully replicated, but our participants often required larger color differences to discriminate between colors. It may reflect uncontrolled sources of variability in crowdsourced experiments, such as variations in the population sample, participant display characteristics (e.g., monitor calibration, brightness, and contrast), or viewing environments.

\parahead{Limitations} The limitations present in this study were similar to those of Experiment~1. Additionally, the original study collected a large sample through LabintheWild~\cite{Reinecke2015LabInTheWild}, whereas we recruited a screened, paid sample two orders of magnitude smaller from Prolific. These differences in recruitment and sample size may also have contributed to the shift in the discrimination threshold.

\section{Conclusion}

\label{sec:discussion}

We replicated two color-discrimination studies 10 years after their original publications, using new implementations and participants. 
We reproduced the core perceptual patterns reported in the original works and then reused them to investigate whether self-reported color-related practice affects color discrimination. This structure follows Quadri~\etals{}~\cite{Quadri2019CantPublishReplication} view that replication studies in visualization can contribute by re-evaluating prior findings and extending them to new questions and settings.
In Experiment~1, discrimination improved with larger marks and color differences, consistent with Szafir's study~\cite{szafir2018modelingColorDifferenceF}. In Experiment~2, we recovered the original right-skewed distribution of discrimination ellipsoid volumes~\cite{reinecke2016enablingDesignersToFores}. 
The results provide evidence that the main findings persist across changes in time, implementation, and participant recruitment.

The key discrepancies from the original were in distinguishing the smallest scatterplot marks (Exp 1), which cover only a few pixels, and in an overall increase in the ellipsoid volumes (Exp 2) on uncontrolled monitors. 
Our studies were conducted 10 years after the originals, during which user displays, hardware, and screen resolutions changed substantially, and our participants were a different population recruited through Prolific rather than the original samples. Hence, the differences we observed may have resulted from the properties of the display conditions, adaptive brightness technologies, or viewing setup rather than from color perception.

We next examined whether self-reported color practices affected color discrimination performance. Across both experiments, we found no significant differences between participants who reported such practices and those who did not. We emphasize that a null result is not evidence of no effect; rather, it suggests that regular color practice might not significantly change the perceptual thresholds evaluated by these tasks. This contrast might be a practical insight into visualization. Visualization designers spend more time working with color than the audiences who interpret their visualizations. If color practice improved fundamental color discrimination, perceptual design guidance applied by designers might not generalize well to typical viewers. Design choices under a visualization designer's control, such as mark size and the color axes used to convey information, are likely to have a greater influence on color discrimination than viewers' color-related experience.

Our findings also highlight opportunities for future replication research. In both replications, the observed thresholds appeared to be more affected by variations in display, adaptive brightness, and viewing settings than by variations in individuals' color practices. Although visualization studies routinely screen participants for characteristics such as color-vision deficiency, they often collect limited information about display hardware or viewing conditions. Our results suggest that more consistent reporting of these factors may improve the reproducibility of color discrimination studies. Since regular color practice had no effect on color discrimination, future research should investigate how color practice influences tasks such as palette construction, multivariate visualization, and other higher-level visualization tasks.

\section{Supplementary Materials}

See 
\href{https://color-perception-replication.netlify.app/}{color-perception-replication.netlify.app/} for stimuli and data, and 
 \href{https://github.com/mcnuttandrew/color-perception-replication}{github.com/mcnuttandrew/color-perception-replication} for code. 

\section{Acknowledgements}

We thank Danielle Szafir, Katharina Reinecke, and David Flatla for their online and emailed materials and guidance.

AI tools (e.g., OpenAI Codex) were used in a limited capacity for statistical analysis and debugging during development. 
All results were manually examined and verified.

\bibliographystyle{abbrv-doi}

\bibliography{template}

@article{Bartram2011WhisperDontSceram,
  author    = {Bartram, Lyn and Stone, Maureen C.},
  doi       = {10.1109/TVCG.2010.237},
  journal   = {IEEE TVCG},
  number    = {10},
  pages     = {1444-1458},
  publisher = {IEEE},
  title     = {Whisper, Don't Scream: Grids and Transparency},
  volume    = {17},
  year      = {2011}
}

@article{borland2007rainbow,
  author    = {Borland, David and Ii, Russell M Taylor},
  doi       = {10.1109/mcg.2007.323435},
  journal   = {IEEE CG\&A},
  number    = {2},
  pages     = {14--17},
  publisher = {IEEE},
  title     = {Rainbow color map (still) considered harmful},
  volume    = {27},
  year      = {2007}
}

@article{cutler2026revisit,
  author    = {Cutler, Zach and Wilburn, Jack and Shrestha, Hilson and Ding, Yiren and Bollen, Brian and Nadib, Khandaker Abrar and He, Tingying and McNutt, Andrew and Harrison, Lane and Lex, Alexander},
  doi       = {10.1109/tvcg.2025.3633896},
  journal   = {IEEE TVCG},
  number    = {1},
  pages     = {13-23},
  publisher = {IEEE},
  title     = {ReVISit 2: A Full Experiment Life Cycle User Study Framework},
  volume    = {32},
  year      = {2026}
}

@article{fider2019differences,
  author    = {Fider, Nicole A and Komarova, Natalia L},
  doi       = {10.1057/s41599-019-0341-7},
  journal   = {Palgrave Communications},
  number    = {1},
  publisher = {Springer},
  title     = {Differences in Color Categorization Manifested by Males and Females: A Quantitative World Color Survey Study},
  volume    = {5},
  year      = {2019}
}

@article{GramazioKaren2017CreatingColorPalettesforInfoVis,
  author    = {Gramazio, Connor C. and Laidlaw, David H. and Schloss, Karen B.},
  doi       = {10.1109/tvcg.2016.2598918},
  issn      = {1941-0506},
  journal   = {IEEE TVCG},
  number    = {01},
  pages     = {521-530},
  publisher = {IEEE},
  title     = {{Colorgorical: Creating Discriminable and Preferable Color Palettes for Information Visualization}},
  volume    = {23},
  year      = {2017}
}

@article{Hall2022ProfessionalDifferences,
  author    = {Hall, Kyle Wm. and Kouroupis, Anthony and Bezerianos, Anastasia and Szafir, Danielle Albers and Collins, Christopher},
  doi       = {10.1109/TVCG.2021.3114805},
  journal   = {IEEE TVCG},
  number    = {1},
  pages     = {654-664},
  publisher = {IEEE},
  title     = {Professional Differences: A Comparative Study of Visualization Task Performance and Spatial Ability Across Disciplines},
  volume    = {28},
  year      = {2022}
}

@inproceedings{healey1996choosing,
  author    = {Healey, C.G.},
  booktitle = {Visualization},
  doi       = {10.1109/VISUAL.1996.568118},
  pages     = {263-270},
  publisher = {IEEE},
  title     = {Choosing effective colours for data visualization},
  year      = {1996}
}

@inproceedings{heer2012color,
  author    = {Heer, Jeffrey and Stone, Maureen},
  booktitle = {ACM SIGCHI},
  doi       = {10.1145/2207676.2208547},
  pages     = {1007--1016},
  publisher = {ACM},
  title     = {Color naming models for color selection, image editing and palette design},
  year      = {2012}
}

@article{horiuchi2024color,
  author    = {Horiuchi, Suzuha and Nagai, Takehiro},
  doi       = {10.1038/s41598-024-60283-4},
  journal   = {Scientific Reports},
  number    = {1},
  pages     = {9615},
  publisher = {Nature Publishing Group UK London},
  title     = {Color discrimination repetition distorts color representations},
  volume    = {14},
  year      = {2024}
}

@inproceedings{Kobayashi2025SeeingIsntBelieving,
  author    = {Kobayashi, Sari and Nakamura, Satoshi},
  booktitle = {Australian Conference on Human-Computer Interaction},
  doi       = {10.1145/3764687.3764709},
  isbn      = {9798400720161},
  numpages  = {13},
  pages     = {150–162},
  publisher = {ACM},
  title     = {Seeing Isn’t Believing: How Visual Illusions Distort Color Selection},
  year      = {2025}
}

@inproceedings{LaceMariah2017HUeBandsandPerception,
  author    = {P. Samuel Quinan and Lace Padilla and Sarah H. Creem-Regehr and Miriah Meyer},
  booktitle = {IEEE InfoVis Posters},
  title     = {Hue Bands and Human Perception: Revisiting the Rainbow},
  year      = {2017}
}

@inproceedings{liu2018somewhere,
  author    = {Liu, Yang and Heer, Jeffrey},
  booktitle = {ACM SIGCHI},
  doi       = {10.1145/3173574.3174172},
  isbn      = {9781450356206},
  numpages  = {12},
  pages     = {1–12},
  publisher = {Association for Computing Machinery},
  series    = {CHI '18},
  title     = {Somewhere over the Rainbow: An Empirical Assessment of Quantitative Colormaps},
  url       = {https://doi.org/10.1145/3173574.3174172},
  year      = {2018}
}

@article{mylonas2014gender,
  author    = {Mylonas, Dimitris and Paramei, Galina V and MacDonald, Lindsay},
  doi       = {10.1075/z.191.15myl},
  journal   = {Colour studies: A broad spectrum},
  pages     = {225--239},
  publisher = {John Benjamins Philadelphia, PA},
  title     = {Gender differences in colour naming},
  year      = {2014}
}

@article{ozgen2002acquisition,
  author    = {{\"O}zgen, Emre and Davies, Ian RL},
  doi       = {10.1037//0096-3445.131.4.477},
  journal   = {Journal of experimental psychology: general},
  number    = {4},
  pages     = {477},
  publisher = {American Psychological Association},
  title     = {Acquisition of categorical color perception: a perceptual learning approach to the linguistic relativity hypothesis},
  volume    = {131},
  year      = {2002}
}

@article{PALAN201822,
  author    = {Stefan Palan and Christian Schitter},
  doi       = {https://doi.org/10.1016/j.jbef.2017.12.004},
  issn      = {2214-6350},
  journal   = {Journal of Behavioral and Experimental Finance},
  pages     = {22-27},
  publisher = {Elsevier BV},
  title     = {Prolific.ac—A Subject Pool for Online Experiments},
  url       = {https://www.sciencedirect.com/science/article/pii/S2214635017300989},
  volume    = {17},
  year      = {2018}
}

@article{Quadri2019CantPublishReplication,
  author  = {Ghulam Jilani Quadri and Paul Rosen},
  doi     = {10.48550/arxiv.arXiv:1908.08893},
  eprint  = {1908.08893},
  journal = {Vis X Vision},
  title   = {You Can't Publish Replication Studies (and How to Anyways)},
  year    = {2019}
}

@article{razali2011power,
  author  = {Razali, Nornadiah Mohd and Wah, Yap Bee},
  isbn    = {978-967-363-157-5},
  journal = {Journal of statistical modeling and analytics},
  number  = {1},
  pages   = {21--33},
  title   = {Power comparisons of shapiro-wilk, kolmogorov-smirnov, lilliefors and anderson-darling tests},
  volume  = {2},
  year    = {2011}
}

@inproceedings{Reinecke2015LabInTheWild,
  author    = {Reinecke, Katharina and Gajos, Krzysztof Z.},
  booktitle = {ACM CSCW},
  doi       = {10.1145/2675133.2675246},
  isbn      = {9781450329224},
  numpages  = {15},
  pages     = {1364–1378},
  publisher = {Association for Computing Machinery},
  series    = {CSCW '15},
  title     = {LabintheWild: Conducting Large-Scale Online Experiments With Uncompensated Samples},
  url       = {https://doi.org/10.1145/2675133.2675246},
  year      = {2015}
}

@inproceedings{reinecke2016enablingDesignersToFores,
  author    = {Reinecke, Katharina and Flatla, David R. and Brooks, Christopher},
  booktitle = {ACM SIGCHI},
  doi       = {10.1145/2858036.2858077},
  isbn      = {9781450333627},
  numpages  = {12},
  pages     = {2693–2704},
  publisher = {ACM},
  title     = {Enabling Designers to Foresee Which Colors Users Cannot See},
  year      = {2016}
}

@article{robson2021EffectOfExpertise,
  author    = {Robson, Samuel G and Tangen, Jason M and Searston, Rachel A},
  doi       = {10.1186/s41235-021-00282-5},
  journal   = {Cognitive Research: Principles and Implications},
  number    = {1},
  pages     = {16},
  publisher = {Springer},
  title     = {The effect of expertise, target usefulness and image structure on visual search},
  volume    = {6},
  year      = {2021}
}

@inproceedings{stone2014engineering,
  author    = {Stone, Maureen and Szafir, Danielle Albers and Setlur, Vidya},
  booktitle = {Color and Imaging Conference},
  doi       = {10.2352/cic.2014.22.1.art00045},
  pages     = {253--258},
  publisher = {Society for Imaging Science and Technology},
  title     = {An engineering model for color difference as a function of size},
  volume    = {22},
  year      = {2014}
}

@article{szafir2014adapting,
  author    = {Danielle Albers Szafir and Maureen Stone and Michael Gleicher},
  doi       = {10.2352/CIC.2014.22.1.art00040},
  journal   = {Color and Imaging Conference},
  number    = {1},
  pages     = {228--228},
  publisher = {Society for Imaging Science & Technology},
  title     = {Adapting Color Difference for Design},
  volume    = {22},
  year      = {2014}
}

@article{szafir2018modelingColorDifferenceF,
  author    = {Szafir, Danielle Albers},
  doi       = {10.1109/tvcg.2017.2744359},
  issn      = {1941-0506},
  journal   = {IEEE TVCG},
  number    = {01},
  pages     = {392-401},
  publisher = {IEEE},
  title     = {{Modeling Color Difference for Visualization Design}},
  volume    = {24},
  year      = {2018}
}

@article{ware2019measuring,
  abstractnote = {Not Available},
  author       = {Ware, Colin and Turton, Terece L. and Bujack, Roxana and Samsel, Francesca and Shrivastava, Piyush and Rogers, David H.},
  doi          = {10.1109/TVCG.2018.2855742},
  journal      = {IEEE TVCG},
  number       = {9},
  pages        = {2777-2790},
  place        = {United States},
  publisher    = {IEEE},
  title        = {Measuring and Modeling the Feature Detection Threshold Functions of Colormaps},
  volume       = {25},
  year         = {2019}
}

@article{ware2023rainbow,
  author    = {Ware, Colin and Stone, Maureen and Szafir, Danielle Albers},
  doi       = {10.1109/mcg.2023.3246111},
  journal   = {IEEE CG\&A},
  number    = {3},
  pages     = {88--93},
  publisher = {IEEE},
  title     = {Rainbow colormaps are not all bad},
  volume    = {43},
  year      = {2023}
}

@article{west2021best,
  title={Best practice in statistics: Use the Welch t-test when testing the difference between two groups},
  author={West, Robert M},
  journal={Annals of clinical biochemistry},
  volume={58},
  number={4},
  pages={267--269},
  year={2021},
  publisher={Sage Publications Sage UK: London, England}
}

@misc{ishihara1951tests,
  title={Tests for colour-blindness},
  author={Ishihara, Shinobu},
  year={1951},
  publisher={Nippon Isho Shuppan Company}
}

@misc{ISO8596_2017,
  organization = {International Organization for Standardization},
  author = {International Organization for Standardization},
  title = {Ophthalmic optics - Visual acuity testing -- Standard and clinical optotypes and their presentation},
  number = {ISO 8596:2017},
  year = {2017},
  address = {Geneva, Switzerland}
}
\clearpage{}
\end{document}